\documentclass[
twocolumn,
superscriptaddress,
amsmath,amssymb,
aps,
prb,
floatfix
]{revtex4-2}

\usepackage[utf8]{inputenc}
\usepackage[T1]{fontenc}
\usepackage{color}
\usepackage[separate-uncertainty=true]{siunitx}
\usepackage{lineno}
\usepackage{graphicx}
\usepackage{amsmath}
\usepackage[colorlinks]{hyperref}

\DeclareSIUnit{\rpm}{rpm}
\DeclareSIUnit{\pixel}{pixel}

\graphicspath{{Figures/}}

\begin{document}

\title{Soil-mimicking microfluidic devices reveal restricted flagellar motility of \textit{Bradyrhizobium diazoefficiens} under microconfinement}

\author{Moniellen Pires Monteiro}
\affiliation{Departamento de F\'isica - Facultad de Ciencias F\'isicas y Matem\'aticas, Universidad de Chile, Beauchef 850, Santiago, Chile}

\author{Juan Pablo Carrillo-Mora}
\affiliation{Departamento de F\'isica - Facultad de Ciencias F\'isicas y Matem\'aticas, Universidad de Chile, Beauchef 850, Santiago, Chile}

\author{Nahuel Guti\'errez}
\affiliation{Facultad de Matem\'atica, Astronom\'ia, F\'isica y Computaci\'on, Universidad Nacional de C\'ordoba, Medina Allende s/n, X5000HUA, Córdoba, Argentina}

\author{Sof\'ia Montagna}
\affiliation{Facultad de Matem\'atica, Astronom\'ia, F\'isica y Computaci\'on, Universidad Nacional de C\'ordoba, Medina Allende s/n, X5000HUA, Córdoba, Argentina}

\author{An\'ibal R. Lodeiro}
\affiliation{Instituto de Biotecnolog\'ia y Biolog\'ia Molecular - Facultad de Ciencias Exactas and CCT-La Plata CONICET, Universidad Nacional de La Plata, B1900, La Plata, Argentina}
\affiliation{Laboratorio de Gen\'etica - Facultad de Ciencias Agrarias y Forestales, Universidad Nacional de La Plata, B1900, La Plata, Argentina}

\author{Mar\'ia Luisa Cordero}
\email{mcordero@ing.uchile.cl}
\affiliation{Departamento de F\'isica - Facultad de Ciencias F\'isicas y Matem\'aticas, Universidad de Chile, Beauchef 850, Santiago, Chile}

\author{V. I. Marconi}
\email{veronica.marconi@unc.edu.ar}
\affiliation{Facultad de Matem\'atica, Astronom\'ia, F\'isica y Computaci\'on, Universidad Nacional de C\'ordoba, Medina Allende s/n, X5000HUA, Córdoba, Argentina}
\affiliation{Instituto de Física E. Gaviola IFEG-CONICET, Medina Allende s/n, X5000HUA, Córdoba, Argentina}

\begin{abstract}
{\it Bradyrhizobium diazoefficiens} is a nitrogen-fixing symbiont of soybean, worldwide used as biofertilizer. This soil bacterium possesses two flagellar systems enabling it to swim in water-saturated soils. However,  the motility in soil pores, which may be crucial for competitiveness for root nodulation, is difficult to predict. To address this gap,  we fabricated microfluidic devices with networks of connected microchannels surrounding grains. In them, we directly visualise bacterial behaviour in transparent geometries mimicking minimalist soils-on-a-chip (SOCs). 
We measured the population velocities and changes of direction for two strains: the wild-type and a mutant with only a subpolar flagellum. A detailed statistical analysis revealed that both strains exhibited reduced speeds and increased changes of direction of 180°, in channels of decreasing cross sectional area, down to a few microns in width. Interestingly, while the wild-type strain displayed faster swimming in unconfined spaces, this advantage was negated in the SOCs with the narrowest microchannels. We employed the measured motility parameters to propose a realistic model and  simulate {\it B. diazoefficiens} confined dynamics being able to reproduce their behaviour, which additionally can be extended enabling further predictions for long time and macro scales. This multidisciplinary work, combining design, microfabrication, microbiology and modelling, offers useful methods to study soil bacteria and may be readily extended to other beneficial/harmful soil species.
\end{abstract}


\maketitle

\section*{INTRODUCTION}
Most bacterial species thriving in soil or aquatic environments display flagella-driven self-propelled motions, which enable them to explore the habitat. Among these environmental microbes, the Hyphomicrobiales order is of paramount importance because it contains most of the N$_2$-fixing species symbiotic with legume plants (hereafter referred to as rhizobia). Despite the large energy consumption required for flagellar assemblage and functioning, flagellar genes were detected in 141 Hyphomicrobiales complete genomes among 150 analysed so far~\cite{Garrido-Sanz2019}, thus highlighting the importance of these appendages for bacterial fitting to the soil environment and interaction with plants.

Inoculation of legume crops with symbiotic rhizobia to profit from N$_2$-fixation has been routinely practised since more than a century ago~\cite{Catroux2001}. Rhizobia are inoculated on seeds or sowing furrows from which they must access the infection targets at the root surfaces to invade the plant tissues and form N$_2$-fixing nodules~\cite{Catroux2001, Basile2021}. Therefore, rhizobia should move in the soil from the inoculation sites to the infection targets either passively dragged by water movement or actively by flagellar motility, although how this transit from the inoculation site to the infection targets is achieved in the soil is under debate since more than 50 years~\cite{Aroney2021}. The soil is a dark, heterogeneous matrix filled with different-sized pores and channels that are narrow and tortuous, wherein rhizobia must move. In addition, the water content is not constant in soil, varying from flooding to several degrees of water deficit in both time and space. Therefore, bacterial movement in the soil cannot be observed directly and has been mostly inferred from indirect observations using motility-altered mutants.

\textit{Bradyrhizobium diazoefficiens} is a symbiotic partner of soybean, a pulse cultivated in more than 120 million hectares worldwide. This bacterium possesses two flagellar systems: a subpolar main system and an inducible lateral system~\cite{Kanbe2007, Althabegoiti2008, Quelas2016, Mongiardini2017, Mengucci2020, Dardis2021}. Recurrent selection in \textit{B. diazoefficiens} allowed obtaining a strain with higher motility, which, once inoculated in soybean, induced significant increases in grain yield at field conditions~\cite{Althabegoiti2008, Lopez-Garcia2009}.

\begin{figure*}[ht]
\includegraphics[width=0.7\textwidth]{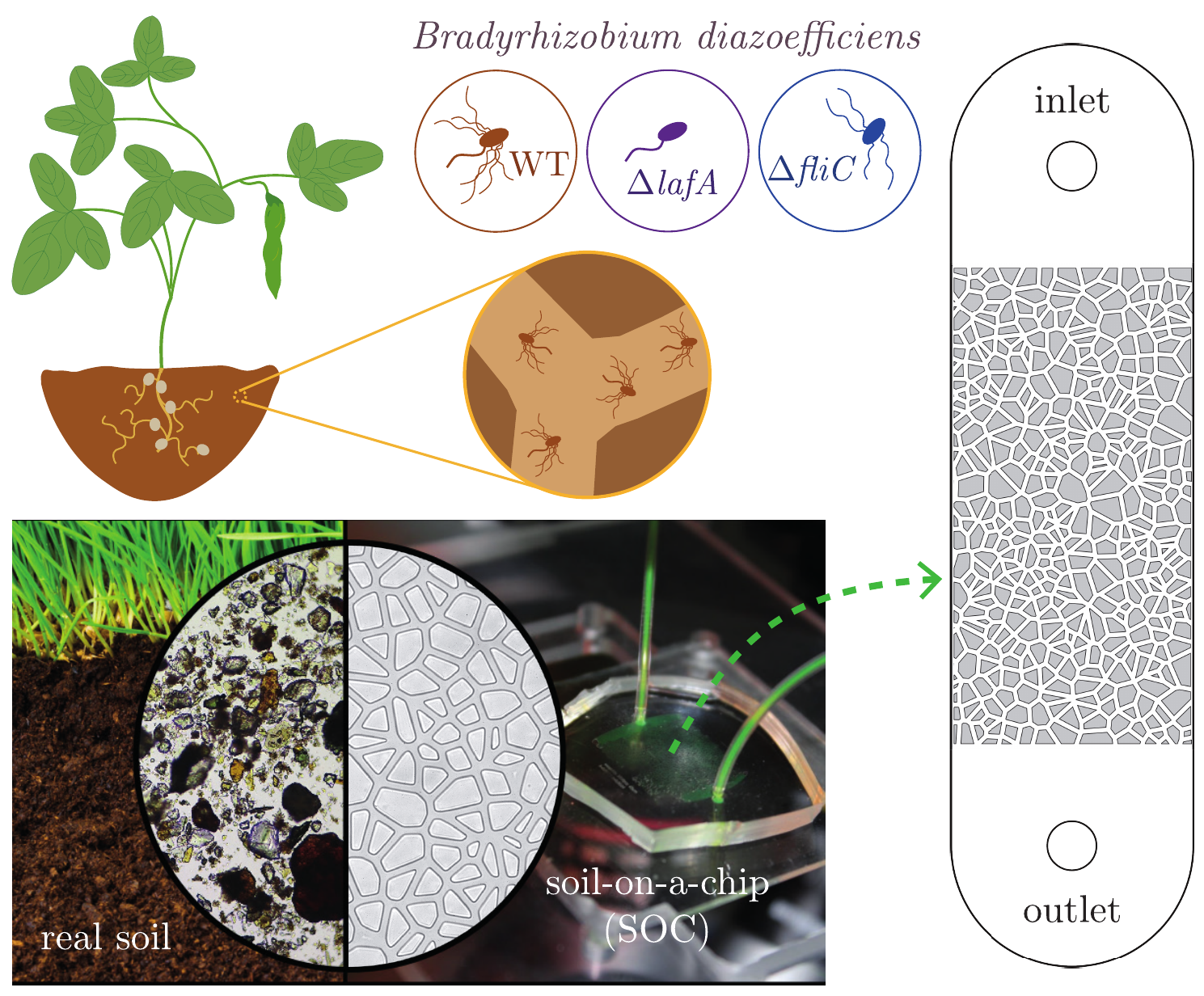}
\caption{\label{fig:Soil-on-a-chip} \textbf{Direct observation of soil bacteria into transparent chambers with microchannels of different widths}. In this scheme, flagellated soil bacteria navigate the soil by swimming through flooded pores between soil particles and grains (left semicircle). To visualize its swimming through a complex geometry with different degrees of confinement, transparent chambers with microchannels of different widths were used (right semicircle, showing a microchannel network of \SI{20}{\micro\meter} channel width). Microfluidic devices, as sketched at the right, were fabricated based on Voronoi tessellations (see Materials and Methods), with channel widths of increasing confinement (\SI{20}{\micro\meter}, \SI{10}{\micro\meter} and \SI{5}{\micro\meter}). Three different strains of \textit{Bradyrhizobium diazoefficiens} were inoculated in the devices, a wild type (WT) with two flagellar systems and two mutants, one lacking the lateral flagellar system ($\Delta$\textit{lafA}) and the other lacking the subpolar one ($\Delta$\textit{fliC}) depicted as light-brown, violet, and blue bacteria, respectively.  Credit for the real soil image (left semicircle): Katelyn Solbakk, Mikroliv~\cite{solbakk2020}.}
\end{figure*}

More detailed studies of \textit{B. diazoefficiens} mutants defective in either the subpolar, the lateral, or both flagellar systems inoculated on soybean plants cultivated in porous medium with different water levels revealed that competition of the non-flagellated mutant against the wild type for colonization and occupation of soybean root nodules is much more restricted at water saturation than at field capacity. Furthermore, the mutant lacking the lateral flagellar system was more competitive than the wild type~\cite{Althabegoiti2011}.

In \textit{B. diazoefficiens}, as in other flagellated bacteria, two distinct modes of flagellar-driven motility have been identified: swimming and swarming (see Ref.~\cite{Aroney2021} and references therein). In liquid environments, swimming is primarily driven by subpolar flagella, although both flagellar systems play a role in motility within semisolid media~\cite{Kanbe2007, Althabegoiti2011, Quelas2016}. Conversely, swarming, which is an associative movement occurring on moist surfaces of semisolid media, required the function of both flagellar systems, with the lateral flagella making a more significant contribution~\cite{Covelli2013}. Additionally, motility assays conducted in soil under varying moisture conditions revealed that the subpolar flagellar system exerted a dominant influence~\cite{Mongiardini2017}. These findings, together with the above-described results on competition for root colonization and nodulation in porous media, suggest that \textit{B. diazoefficiens} predominantly uses swimming to navigate water-saturated microchannels in the soil. In contrast, swarming motility appears to be dispensable for movement in soil and seems not essential for effective root colonization or nodulation of soybeans.

Swimming of \textit{B. diazoefficiens} was studied and characterized by direct observations of living cells under the microscope, which allowed measuring important parameters associated with each flagellar system, such as swimming speed, angle of direction, and frequencies of changes of direction, flagellar rotation sense, etc.~\cite{Kanbe2007, Althabegoiti2011, Quelas2016}. However, the results of these studies, which were performed in homogeneous liquid environments, are difficult to extrapolate to the real situation in the soil.

Developments in microfluidics, which open doors to microbiology~\cite{Hol2014, Stanley2016, Massalha2017, Tan2020} and its applications~\cite{Nosrati2017, Gharib2022}, have allowed us to prepare transparent and biocompatible microscopic chambers containing a network of connected channels resembling the soil grain complex structure in a very simplified manner, and directly observe bacteria swimming within channels of different widths surrounding grains of different sizes (Fig.~\ref{fig:Soil-on-a-chip}). Using these tools, we were able to directly observe, quantify, model and compare the swimming of the \textit{B. diazoefficiens} USDA 110 wild type (WT) and derived mutants devoid of subpolar ($\Delta$\textit{fliC}) or lateral ($\Delta$\textit{lafA}) flagellar filaments under microconfinement in conditions alike a flooded ideal soil, and extend the previously obtained information from direct observations in homogeneous liquid as well as indirect observations in agar or porous media.

\section*{RESULTS}
\subsection*{Reduction of microbial swimming speed under microconfinement}

We developed microfluidic devices to mimic the intricate geometry of flooded pores in soil. However, it must be remembered that these devices are intended as approximations to simulate only the pore structure of soils. Real soils are more complex, highly heterogeneous structures, formed by grains whose size can range from micrometres to a few millimetres, but, in addition, soil channels may contain surfaces covered by organic matter to different extents, ionized clays, pH gradients, etc. In our devices, we fabricated grains of sizes between \SI{5}{\micro\meter} and \SI{100}{\micro\meter}, which represent silty and fine sandy soils. As grains reduce in size, so do the pores between the grains. Thus, our devices had different degrees of confinement, which was quantified by the width $w_i$ of the channels where bacteria navigate (see Methods). Three different devices were fabricated with increasingly narrower channels, $w_1 = \SI{20}{\micro\meter}, w_2 = \SI{10}{\micro\meter}$, and $w_3 = \SI{5}{\micro\meter}$. In this way, we increment the degree of confinement by progressively halving the width of the channels. As a reference to the dimension of the degree of confinement that the bacterial cells will experience, data in the literature indicate that the typical body length of a \textit{B. diazoefficiens} cell is about \SI{1}{\micro\meter} to \SI{2.5}{\micro\meter} at mid-log phase of growth, and about \SI{10}{\micro\meter} when considering the complete cell including its flagella~\cite{Quelas2016, Montagna2018, Cogo2019, Mengucci2020, Medici2024}. Hence, the effect of walls is expected to increase as confinement becomes tighter, for example modifying the individual bacteria swimming strategies and the collective population transport properties (diffusion phenomena).

Using bright field microscopy and digital imaging, we recorded and tracked the trajectory of thousands of bacteria of the three mutants (see Methods) swimming in the microfluidic devices. We observed that the vast majority of $\Delta$\textit{fliC} mutants (strain with lateral flagella only) moved irregularly, as trembling in a given position, without noticeable translocation (see Supplementary Movie 1), while only a very low percentage of cells were clearly swimming persistently, whereby we decided not to continue with the analysis of this mutant. 

\begin{figure*}[p]
\centering
\includegraphics[width=0.85\linewidth]{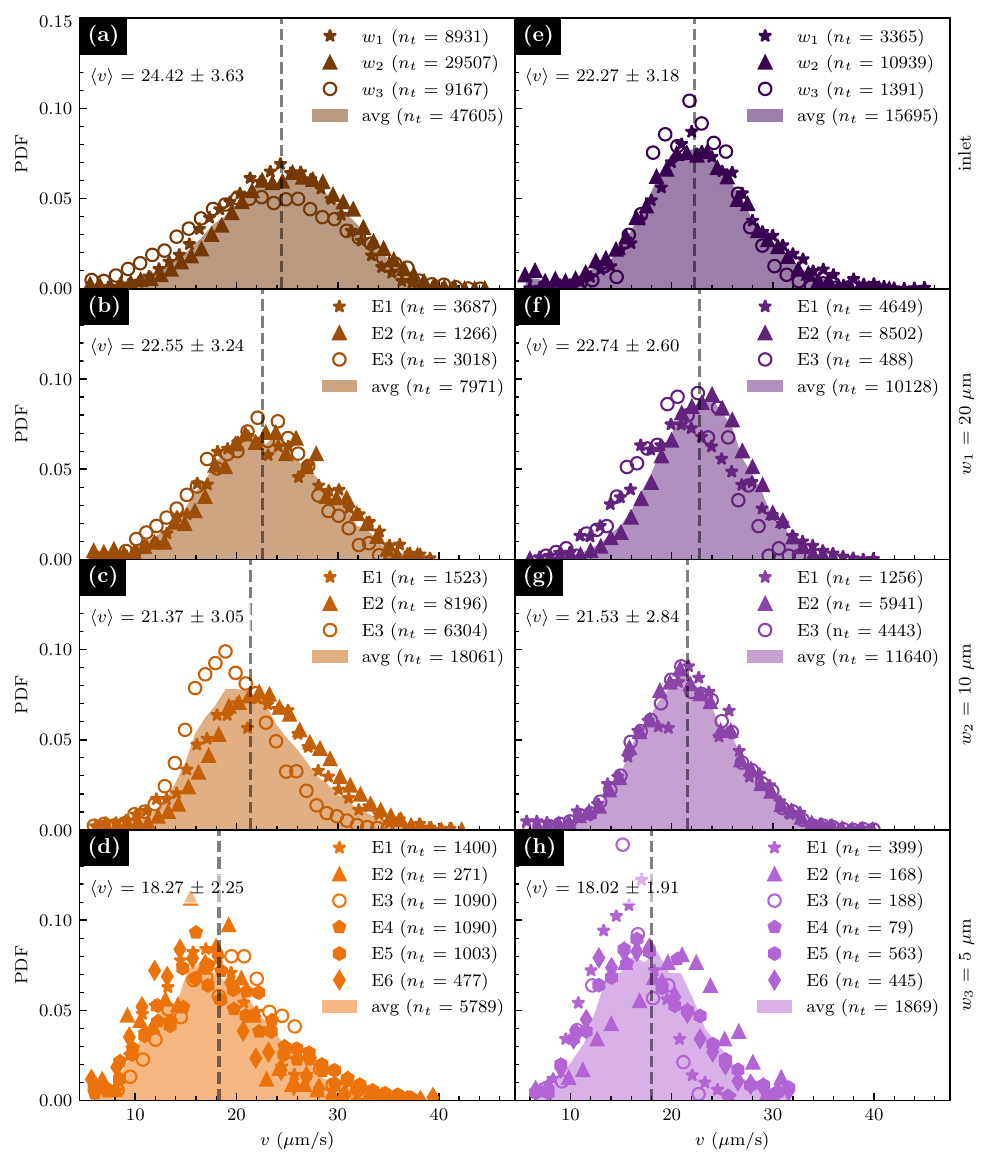}
\caption{\label{fig:SpeedDistribution} \textbf{Speed distributions of bacteria in microfluidic devices.} Speed probability density function (PDF) of WT [(a)--(d)] and $\Delta$\textit{lafA} [(e)--(h)] \textit{B. diazoefficiens} strains are shown for measurements into the inoculation site at the inlet [panels (a) and (e)], or a porous geometry with channels of width $w_1 = \SI{20}{\micro\meter}$ [(b) and (f)], $w_2 = \SI{10}{\micro\meter}$ [(c) and (g)], and $w_3 = \SI{5}{\micro\meter}$ [(d) and (h)]. Symbols represent experimental distributions from 3 to 6 biological replicas (E$_i$, with $i=1, \ldots, 6$), with a number of observed tracks $n_t$ indicated for each case in the legends. Solid light-coloured backgrounds represent the average of all experiments (avg). Mean speed values $\pm$ error (in units of \si{\micro\meter/\second}) are given in all panels and indicated with a dashed vertical line.}
\end{figure*}

\begin{figure*}[htb]
\centering
\includegraphics{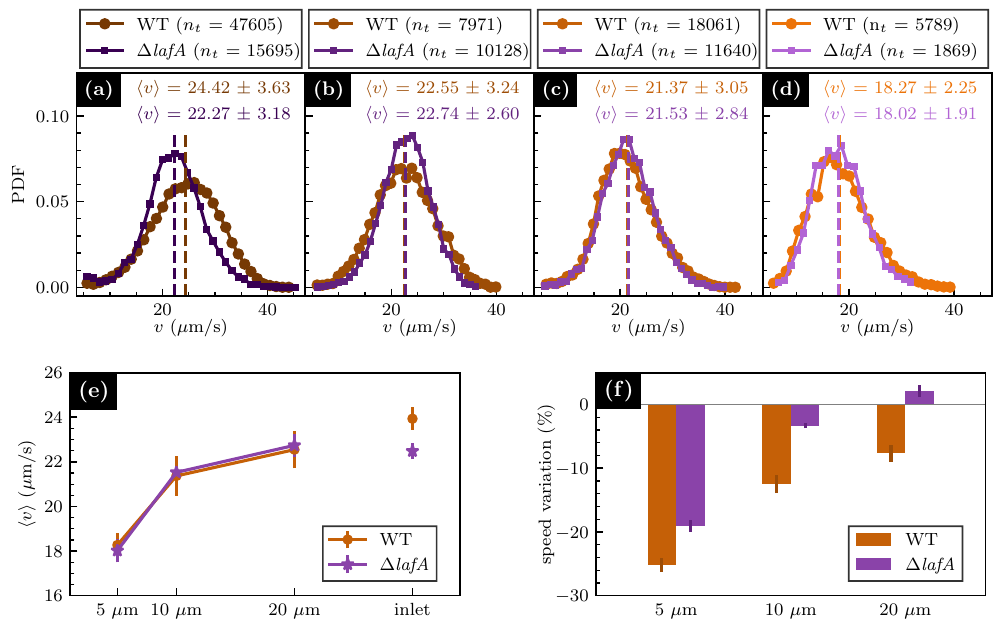}
\caption{\label{fig:MeanSpeed} \textbf{Average speed and speed variation ratio in microfluidic devices with different degrees of confinement.} Speed PDF for WT and $\Delta$\textit{lafA} mutant in the inoculation chamber (a) and in microfluidic devices with different degrees of confinement: $w_1=\SI{20}{\micro\meter}$ (b), $w_2=\SI{10}{\micro\meter}$ (c), and $w_3=\SI{5}{\micro\meter}$ (d). (e) Bacteria average speed in the microfluidic devices as a function of the channel width. (f) SVR for each strain as a function of the channel width (see Eq.~\eqref{eq:svr}). Vertical bars in both cases indicate the data corresponding errors.}
\end{figure*}

Fig.~\ref{fig:SpeedDistribution} shows the probability density function (PDF) of the bacteria speed for the WT and $\Delta$\textit{lafA} strains with both flagellar systems and with a subpolar flagellum only, respectively, swimming in the three devices. All experiments were repeated with at least three biological replicas to assess data reproducibility and to obtain good statistics. In each experiment, bacteria trajectories were also measured in the inlet compartments (see Fig.~\ref{fig:Soil-on-a-chip}), where bacteria swim in a quasi-two-dimensional chamber without grains or obstacles. As a reference, the speed distributions obtained for each strain in these inlet regions are shown in the first row of Fig.~\ref{fig:SpeedDistribution}.

In general, differences in the speed PDFs between biological replicas are minor and reflect some diversity among different cultures. However, when all data are combined, the resultant speed PDF for each population is smooth, reflecting the high-quality statistics; they were obtained with thousands of cell tracks lasting more than \SI{0.5}{\second} each. This time of tracking is difficult to achieve in these kinds of setups. It is roughly equivalent to the translocation of 12-15 body cells on average, a time-lapse during which bacteria swim in focus. In the inlets, the speed PDFs for both strains are single-peaked and symmetric, with most of the cells swimming at speeds close to the maximum of the distribution and a small population of bacteria swimming at speeds smaller than \SI{10}{\micro\meter/\second} and larger than \SI{35}{\micro\meter/\second}. As the width of the channels decreases, the observed behaviour is similar for both strains: the speed distributions remain single-peaked, but the swimming speeds become progressively slower, which is manifested by a systematic shifting of the distributions toward the left. In the case of the WT, this shifting is already evident in the channels with width \SI{20}{\micro\meter} and continues shifting towards smaller speeds as the channel width decreases. In the case of the $\Delta$\textit{lafA}, the speed distribution in the widest channels does not differ appreciably with respect to the inlet. However, a systematic decrease in mean speed can be seen in the networks with channel width \SI{10}{\micro\meter} and \SI{5}{\micro\meter}.

We compared the speed PDFs of both strains in the different geometries in Fig.~\ref{fig:MeanSpeed}(a)--(d). In the inoculation chambers (Fig.~\ref{fig:MeanSpeed}(a)), the speed distributions of the WT and the mutant strain are appreciably different. That for the WT is displaced to higher values than the mutant's speed distribution. In quantitative terms, the mean speed of the WT in the inlet is \SI{24.42 \pm 3.63}{\micro\meter/\second}, and that of the single-flagellated $\Delta$\textit{lafA} is \SI{22.27 \pm 3.18}{\micro\meter/\second}. That corresponds to a mean speed difference of almost \SI{10}{\percent}. This behaviour indicates that, on average, the presence of the two flagellar systems propels the WT bacteria at a higher speed. While in the $\Delta$\textit{fliC} mutant, the lateral flagella produce irregular and poor swimming, a complex cooperative behaviour between the two flagellar systems appears to be at work in the case of the WT, allowing them to swim faster than the $\Delta$\textit{lafA} mutant. The speed distribution is also wider for the WT than for the $\Delta$\textit{lafA} strain in the inlet. This could be explained by natural variations in the number and length of lateral flagella, which causes population diversity in the case of the WT. In comparison, the $\Delta$\textit{lafA} mutant, only possessing a single flagellar system, has fewer degrees of freedom associated with the flagellar motor system, thus resulting in a more homogeneous population and a more concentrated speed distribution.

As the available space to swim is reduced, the mean speed of both strains decreases, as explained previously. In the microfluidic devices, the average swimming speed coincides almost exactly for both strains in all channel widths, including the widest one, even though, on average, the WT strain swims faster than the mutant in unconfined space (Fig.~\ref{fig:MeanSpeed}(b)--(c)). Moreover, as confinement increases, the speed distributions of both strains become more alike. In the channels of width \SI{20}{\micro\meter}, the speed distribution of the WT is still wider than that of the $\Delta$\textit{lafA}, but in the channels of width \SI{10}{\micro\meter} and \SI{5}{\micro\meter}, the speed PDFs of both strains are almost identical. The only notable difference between both strains is that the speed distributions for the WT in the microfluidic devices become asymmetric, systematically showing longer tails at large speeds than the $\Delta$\textit{lafA}. This reveals that, although the average speed of the overall population decreases in both strains, and the population diversity for the WT is reduced, a small but non-negligible fraction of WT bacteria maintains high swimming speeds, even in the most confined geometries. On the contrary, the speed PDFs for the $\Delta$\textit{lafA} remain symmetric, showing that the overall population of bacteria decreases its swimming speed.

The average speed for each channel width is plotted in Fig.~\ref{fig:MeanSpeed}(e). The fall of mean speed is evident in the microfluidic device with narrower channels. To quantify this speed reduction with respect to the mean speed in the inlet chamber, we define the speed variation ratio (SVR) as
\begin{equation}
\textrm{SVR} = \frac{\langle v \rangle - \langle v \rangle_\textrm{inlet}}{\langle v \rangle_\textrm{inlet}}\times 100\%,
\label{eq:svr}
\end{equation}
where $\langle v \rangle_\textrm{inlet}$ is the average speed in the inlet for each strain. The SVR is presented in Fig.~\ref{fig:MeanSpeed}(f). In the widest channels, the SVR is about \SI{-7.5}{\percent} for the WT and \SI{2}{\percent} for the mutant. As the confinement increases, the SVR falls for both strains, reaching \SI{-12.5}{\percent} and \SI{-25}{\percent} for the WT in the channels with width \SI{10}{\micro\meter} and \SI{5}{\micro\meter}, respectively. The corresponding SVR for the $\Delta$\textit{lafA} strain in the same confinements are \SI{-3}{\percent} and \SI{-19}{\percent}. In this case, the speed reductions are lower than in the WT because the reference value in the inlet is lower for the mutant, but the average values in the microchannels are very similar for both strains.

Overall, the results shown above indicate that confinement affects the swimming behaviour of both strains. However, it appears to affect each strain differently, as can be seen when comparing the speed distributions of each strain in the microchannels to their respective distributions in the inlet. For the mutant inside the microchannels, although the mean speed decreases systematically with decreasing channel width, the shape of the speed distribution remains almost unaltered; the distribution's height, width, and symmetry remain approximately constant. For the WT strain, conversely, the speed distributions become narrower, taller, and asymmetric in the confining microfluidic devices, and both in the mean speed and in the shape of the speed PDF, its behaviour becomes similar to the $\Delta$\textit{lafA} mutant. The cooperative behaviour between the two flagellar systems observed for WT swimming in the inlets appears to be lost, on average, when they swim in the microchannels.

\subsection*{Swimming strategies of \textit{B. diazoefficiens} in microchannels}

\begin{figure*}[p]
\centering
\includegraphics[width=0.95\textwidth]{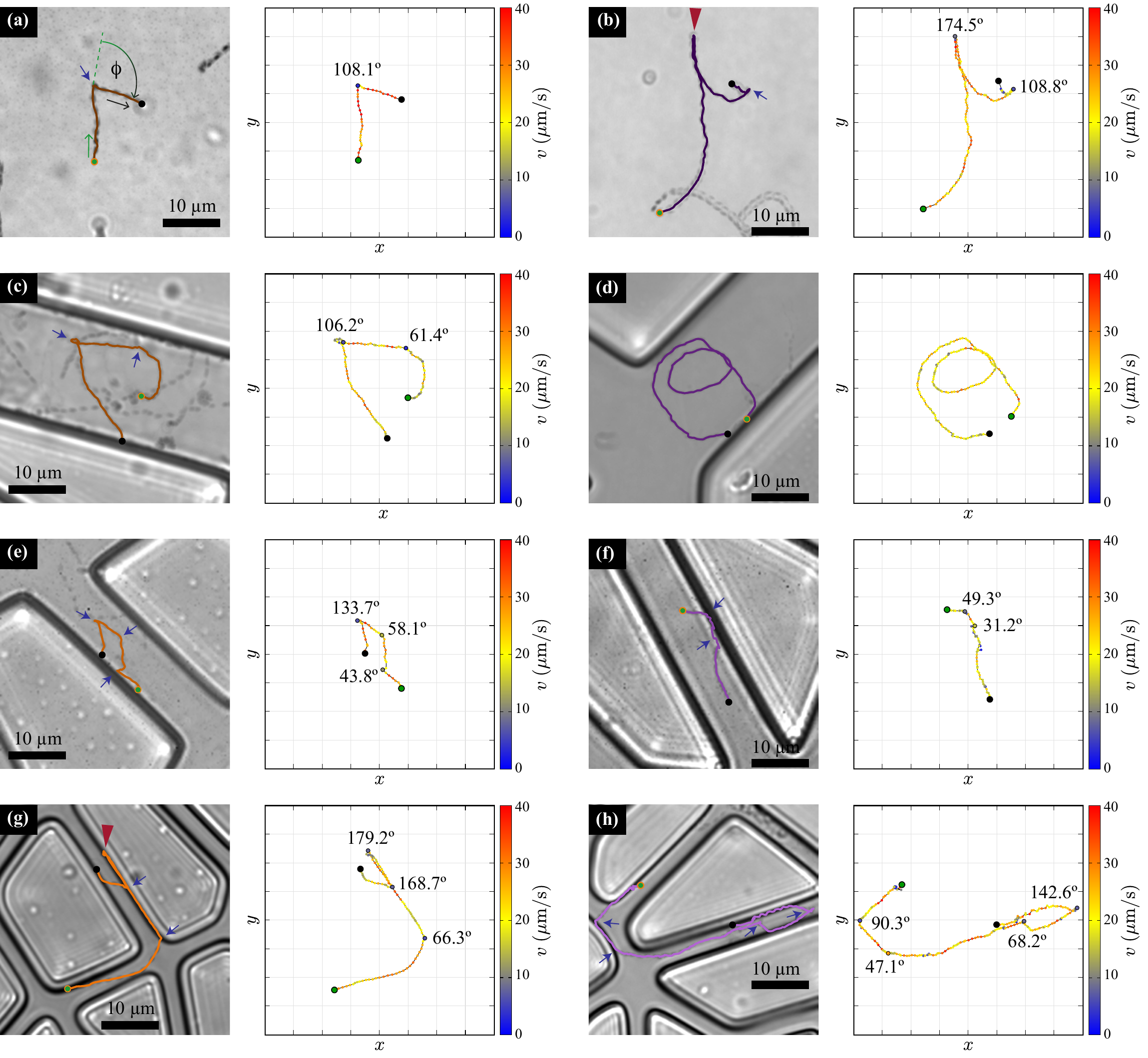}
\caption{\label{fig:Trajectories} \textbf{Swimming behaviour of \textit{B. diazoefficiens} in unconfined space and in microchannels.} Examples of trajectories followed by WT (left) and $\Delta$\textit{lafA} (right) at the unconfined inlets [(a) and (b)] and in the microchannels with increasing confinement [$w_1 = \SI{20}{\micro\meter}$ (c) and (d); $w_2 = \SI{10}{\micro\meter}$ (e) and (f); and $w_3 = \SI{5}{\micro\meter}$ (g) and (h)]. For each case, the trajectory is shown at the left and at the right, the bacterium speed is presented along their path in colour scale, with the angle values for each detected change of direction, $\Phi$. Several snapshots are overlaid to show the trajectory of the bacteria for the time span of the tracks, and only one bacterium track is shown. The green and black dots mark, respectively, the start and end of the track. The blue arrows and red arrowheads mark changes of direction (CHD) and run and reverses (RR), respectively. In (a), the reorientation event is marked with the definition of the reorientation angle, $\Phi$. Videos of the trajectories are available as Supplementary Movies 2-9.}
\end{figure*}

An important component of the behaviour of swimming bacteria is their specific strategies to rectify their direction of movement by reorienting during swimming. For soil bacteria, this behaviour should be highly dependent on the porous media they navigate to live and survive. For this reason, we characterized the features of bacteria paths in all regions of the microfluidic devices.

Fig.~\ref{fig:Trajectories} shows different trajectory examples in the inlets and the microchannels for the WT and the $\Delta$\textit{lafA} strains. These bacteria commonly reorient during their swimming. The angle $\Phi$ quantifies the change of direction (CHD) events. We define $\Phi$ as the positive difference in orientation angle between the final and initial direction of motion, as sketched in Fig.~\ref{fig:Trajectories}(a). When $\Phi$ is small, $\Phi \in [\SI{15}{\degree}, \SI{35}{\degree}]$, the motion is considered highly persistent, and bacteria move essentially forward. On the contrary, CHDs with $\Phi > \SI{90}{\degree}$ contribute to the backward motion of the bacteria.

Fig.~\ref{fig:Trajectories}(a) presents a single, evident CHD for a WT bacterium to show our definition of CHD and $\Phi$. In this case, $\Phi = \SI{108.1}{\degree}$. The track starts with the green dot and finishes with the black one. The blue arrow points at the reorientation event. The contiguous panel evidences the speed along the path, which fluctuates between \SI{20}{\micro\meter/\second} and \SI{40}{\micro\meter/\second}, with an important fall to less than \SI{5}{\micro\meter/\second} at the CHD.

\begin{figure*}[ht]
\centering
\includegraphics[width=\textwidth]{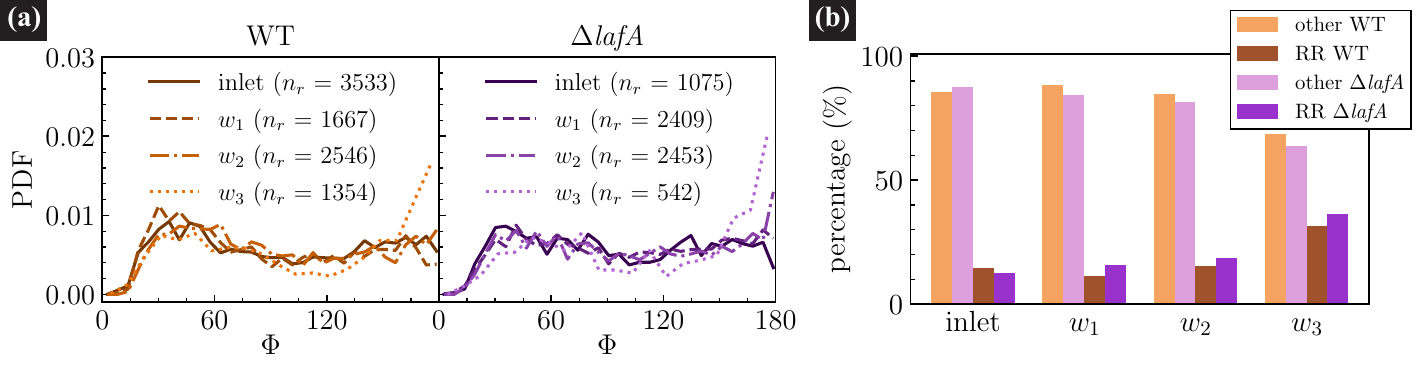}
\caption{\label{fig:AngleDistributions} \textbf{Changes of Directions of \textit{B. diazoefficiens} measured in the microfluidic devices.} (a) Reorientation angle PDF of WT (left) and $\Delta$\textit{lafA} (right) in the inoculation chambers and in the different confined networks with channels width  $w_1 = \SI{20}{\micro\meter}, w_2 = \SI{10}{\micro\meter}$, and $w_3 = \SI{5}{\micro\meter}$. (b) Classification of CHDs in the inlets and in the different microchannels, according to their angle: RR for $\Phi > \SI{160}{\degree}$, and ``other'' otherwise.}
\end{figure*}

One example of bacteria trajectory with CHDs for the $\Delta$\textit{lafA} strain in the inlet, along with the corresponding bacteria speed, is shown in Fig.~\ref{fig:Trajectories}(b). A remarkable behaviour is the common existence of CHDs with an angle $\Phi \approx \SI{180}{\degree}$. These events occur in both strains and are reminiscent of the ``Run and Reverse'' (RR) behaviour exhibited by some monotrichous bacteria under certain circumstances, such as \textit{Vibrio alginolyticus}~\cite{Xie2011, Son2013} and \textit{Caulobacter crescentus}~\cite{Morse2016}. Examples of RR events are marked with a red arrowhead in Fig.~\ref{fig:Trajectories}. Close inspection of the videos suggests that RRs occur when the whole body of the bacterium turns in \SI{180}{\degree} instead of reversing the swimming direction with the body in the same orientation (see videos in Supplementary Material). This implies that the reversal is not due to the flagella reversing their spinning direction. Due to noise, the RR angle is not necessarily equal to \SI{180}{\degree}. It is difficult to associate a threshold value for $\Phi$ to characterize an RR event; however, by inspection of the videos, we have decided to classify all CHDs with $\Phi > \SI{160}{\degree}$ as RRs. In the microchannel networks, bacteria interact with bottom-top surfaces and vertical boundaries. In the case with $w_1 = \SI{20}{\micro\meter}$, bacteria exhibit large trajectories and can perform more than one CHD in the channels before contacting the boundaries. Fig.~\ref{fig:Trajectories}(c) shows one example trajectory of a WT bacterium swimming inside a channel and performing two CHDs before hitting the wall. In Fig.~\ref{fig:Trajectories}(d), one example trajectory of a $\Delta$\textit{lafA} cell can be observed. In this case, the swimmer circles twice between two close vertical walls without hitting them.

In the microfluidic device with $w_2 = \SI{10}{\micro\meter}$, the interaction with boundaries becomes more common. Still, bacteria can swim in the plane for long trajectories and have CHDs before touching a vertical wall. Figures~\ref{fig:Trajectories}(e) and (f) show, respectively, trajectories of a WT and a $\Delta$\textit{lafA} in the microfluidic device with width \SI{10}{\micro\meter}. In both cases, the bacteria swim near the solid boundary for a short distance of approximately \SI{5}{\micro\meter} before a CHD reorients them into the microchannel. We commonly observe this behaviour with in-plane path segments in which a bacterium swims close to the walls for a fraction of a second and then swims again into the microchannel, away from the wall.

In the network with microchannel width $w_3 = \SI{5}{\micro\meter}$, bacteria show qualitatively different behaviours. An example of a WT bacterium swimming in this device is presented in Fig.~\ref{fig:Trajectories}(g). The path presents a relatively long segment before a collision with a solid boundary. After hitting the wall, the bacterium aligns with it. Although the solid boundaries appear dark as well as bacteria, post-processing of the images allowed us to detect the cell swimming for a few seconds very near the wall, even after an RR. Eventually, another CHD causes the cell to cross the channel and align to the opposite wall. Similarly, the trajectory of a $\Delta$\textit{lafA} cell, shown in Fig.~\ref{fig:Trajectories}(h), presents segments in which the cell swims parallel and very close to the walls, followed by short segments in which the bacterium crosses from one wall to the opposite one. Most CHDs in this device occur when the bacteria either encounter a wall or while swimming aligned to a wall. A minority of CHDs are detected when the bacterium is within one channel, away from the borders.

The distributions of reorientation angle $\Phi$, in the inlets and in the network of grains and channels, for both strains are shown in Fig.~\ref{fig:AngleDistributions}(a). In the inlets, for both strains, the probability for a CHD to have an angle $\Phi \in [\SI{30}{\degree}, \SI{180}{\degree}]$ is quite uniform, with only a small increase of approximately \SI{5}{\percent} in probability around $\Phi = \SI{40}{\degree}$ and $\Phi = \SI{180}{\degree}$, as compared to the probability at the centre of the distribution. Note that small angles of change of direction  $\Phi < \SI{30}{\degree}$ are difficult to detect, so their probability is underestimated. The distribution does not change in the coarser microdevice with $w_1 = \SI{20}{\micro\meter}$. In the microdevice with $w_2 = \SI{10}{\micro\meter}$, the probability distribution for the WT remains similar, but in the case of the mutant, there is an increase of probability for CHDs with a large angle, $\Phi \approx \SI{180}{\degree}$. The trend is confirmed in the microchannels of width $w_3 = \SI{5}{\micro\meter}$, where there is an enlarged probability for $\Phi > \SI{160}{\degree}$ in both strains. This is consistent with the observed trajectories in this device, where RRs were common for bacteria swimming close to solid walls.

Based on the threshold value stated above, we classified CHDs with $\Phi > \SI{160}{\degree}$ as RR and the rest as ``other''. As shown in Fig.~\ref{fig:AngleDistributions}(b), RRs amount to approximately \SI{15}{\ per cent} of all CHDs for both strains in the inlets. The percentage of RR changes is marginal in the two widest microchannels, but interestingly, it shows a significant increase in the mimicked ideal soil with narrowest microchannels with $w_3 = \SI{5}{\micro\meter}$. There, approximately, \SI{32}{\percent} of change of directions for the WT and \SI{37}{\percent} for the single-flagellated $\Delta$\textit{lafA} correspond to RR. These results confirm that no apparent differences between strains exist in what concerns the swimming strategies of \textit{B. diazoefficiens}, an interesting insight regarding the two flagellar systems' functionality under strong microconfinement. Instead, the increasing influence of channel walls when confinement increases guides bacteria into swimming along walls, thus increasing the proportion of RR for both strains.

\subsection*{Simulations of \textit{B. diazoefficiens} in simple soil-mimicking microfluidic devices.}

\begin{figure*}[p]
\centering
\includegraphics[width=\textwidth]{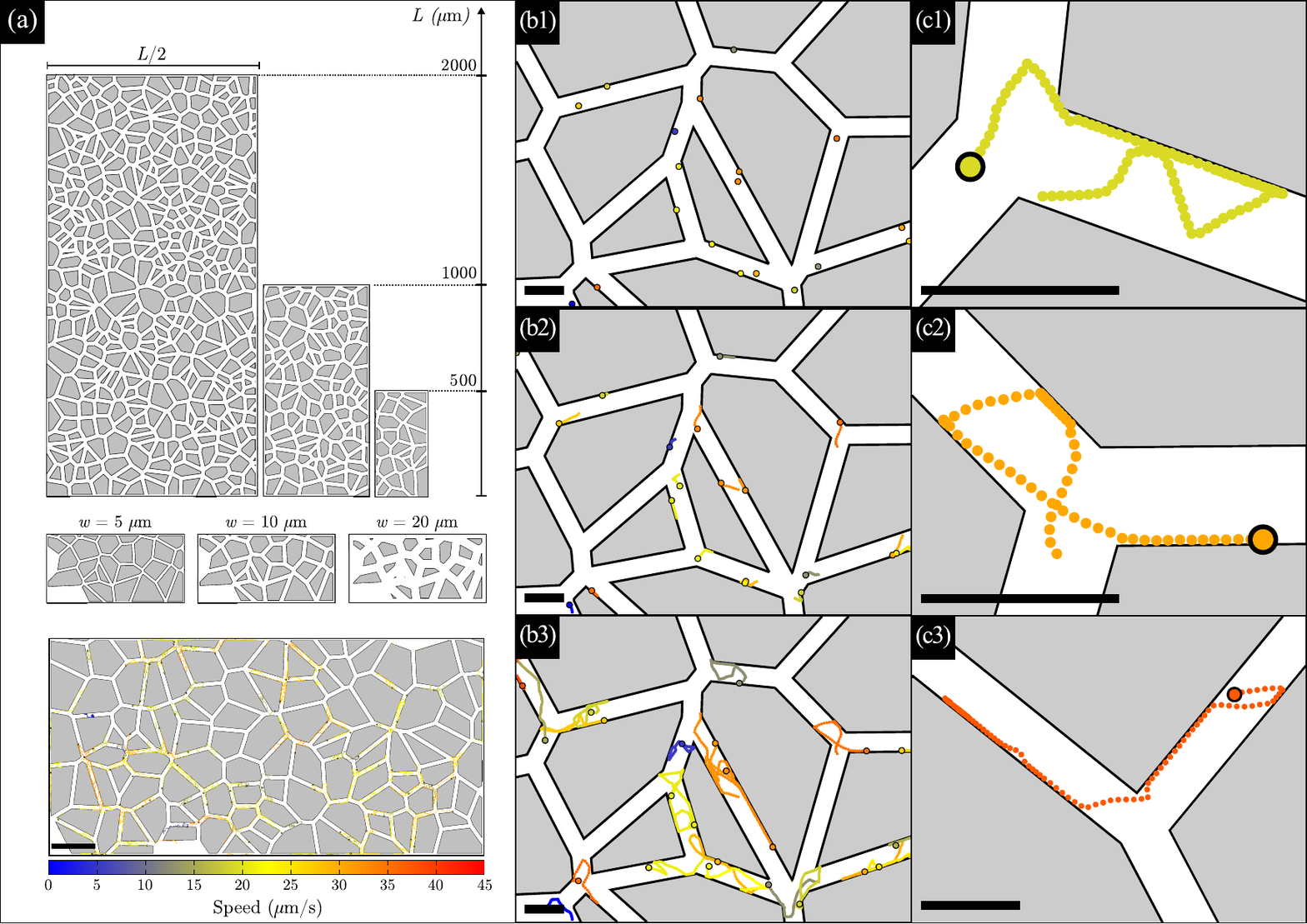}
\caption{\label{fig:Simulations} 
\textbf{Simulated \textit{B. diazoefficiens} swimming behaviour in  microdevices mimicking a porous media with $\SI{50}{\micro\meter}$ mean size grains.} (a) Designed arrays composed of islands and channels, with total area $L/2 \times L$. In the second row, an array with $L = \SI{500}{\micro\meter}$ and three channel widths is shown: $w = 5, 10, 20$ \si{\micro\meter} (left to right respectively). In the lower panel, a few bacteria trajectories are drawn inside a complete array with $L = \SI{1000}{\micro\meter}$, $w= \SI{10}{\micro\meter}$. Scale bar: $\SI{100}{\micro\meter}$. The colour of the tracks corresponds to the speed of each simulated bacterium (proportional to their motor force of self-propulsion), according to the speed colour scale at the bottom. The same colour scale is used in all panels. (b1) Initial positions, at $t = 0$. Dots representing bacteria are enlarged for visualization. (b2) Starting bacteria trajectories until $t = \SI{0.5}{\second}$. The black circles represent the final positions. (b3) Simulated tracks of duration \SI{3.5}{\second}. Most of the bacteria have performed several changes of direction into the middle of the channels or due to interactions. Note the swimming along walls, with some straight paths of more than $\SI{30}{\micro\meter}$. (c1) to (c3) show three examples of simulated tracks of duration \SI{5.5}{\second}, \SI{5}{\second}, and \SI{3.75}{\second}, respectively. Scale bars in (b) and (c) panels correspond to $\SI{20}{\micro\meter}$.}
\end{figure*}

Thanks to the detailed motility parameters measured above for both strains of \textit{B. diazoefficiens}, their motion in confined geometries can be modelled phenomenologically and simulated accurately. Calculations were performed as realistically as possible, imitating both the exact microdevices design, experimental setup and the self-propulsion of each strain (see details in Methods Section). Thus, the numerical simulations allowed reaching length and time scales that are unavailable to experiments in microfluidic devices.

We show in Fig.~\ref{fig:Simulations} examples of the arrays of grains used in the numerical calculations. Planar geometries exactly analogous to the experimental designs are used and shown in panel (a). In simulations, the global size of the microdevice can be varied further to perform scaling studies, but according to the experiments, the simulation area was kept with an aspect ratio of 2:1 (total length:total width). The microchannel width surrounding the impenetrable islands is fixed in each simulation, as a first simplified network mimicking soils, with width ranging between $w = \SI{5}{\micro\meter}$ and $w = \SI{20}{\micro\meter}$ as shown in the second row of Fig.~\ref{fig:Simulations}(a).

Bacteria are located randomly inside the microchannels to start the simulations with the distribution of speeds that imitate the measured speeds, Fig.~\ref{fig:SpeedDistribution}(b)--(d),(f)--(h). The model proposed, using physical laws, is able to predict the positions at all times. To show its capacity at short and large time scales, snapshots of the cells' positions along time are shown with tracks from short up to long duration equal to the measured times in the experiments. A complete track of them into the simulated microchannels is shown in the lower part of Fig.~\ref{fig:Simulations}(a), with the mean speed of each bacterium represented in colour. To improve visualization, only a fraction of the simulated bacteria is drawn. A zoomed-in region is shown in Fig.~\ref{fig:Simulations}(b1-b3). In Fig.~\ref{fig:Simulations}(b1), the initial positions are sketched, and in Fig.~\ref{fig:Simulations}(b2), the trajectories during the first steps simulated are shown. Specifically, at a time $t = \SI{0.5}{\second}$, most of the simulated bacteria have not performed any CHD yet, and mostly straight paths are observed of length proportional to the bacterial speed. In Fig.~\ref{fig:Simulations}(b3), longer tracks of more than three seconds are presented to visualize more details. As observed experimentally, several CHDs and interactions with the walls are observed. In contrast to experimental tracks, numerical simulations have the advantage of not losing any details close to walls or due to the swimming out of the focal plane, resulting in a complementary tool of analysis. The CHDs are seen both close to boundaries, in the middle of the microchannels, or at the microchannel crossings. In addition, long tracks along walls can be detected with mean velocities of \SI{30}{\micro\meter/\second} for more than \SI{30}{\micro\meter}.

Zooming in even more, several specific tracks are presented in Fig.~\ref{fig:Simulations}(c1) to (c3). Fig.~\ref{fig:Simulations}(c1) presents a single cell trajectory of total duration \SI{5.5}{\second}, where a clear RR of \SI{180}{\degree} occurs along the wall. In Fig.~\ref{fig:Simulations}(c2), the bacterium moves from wall to wall, following a track, at first quite circular and later crossing more straightly. Finally, Fig.~\ref{fig:Simulations}(c3) shows a single long track that lasts $\SI{3.75}{\second}$. Note the longer track in a shorter time due to its larger velocity than the previous two examples. A long path following almost a full side of the grain is observed at the starting point, with a short detaching from the wall to come back to follow the wall. Later, a small CHD produces a channel crossing. A large CHD is shown at the end of the detection, very close to an RR. In brief, our model and simulations can reproduce the wide variety and complexity of the swimming behaviour, in good agreement with that observed experimentally under confinement, as the examples shown in Fig.~\ref{fig:Trajectories}, and it can help and be complementary in details hard to be observed in experiments under ultra confinement. The observed agreement is a base step of the modelling and suggests confidence to go further using the predictive model in larger scales and more complicated applications.

\section*{DISCUSSION}

Dual flagellar systems are relatively rare in motile bacterial species. Since the production of the motor and flagellar structures requires an important amount of energy and resources, it has been proposed that these systems have been selected as evolutionary advantages to fit in the ecological niche of each species, fulfilling different and specific roles in motility. In what concerns \textit{B. diazoefficiens}, different functionalities of its two flagellar systems have been proposed in the past, but direct evidence is still lacking to validate current hypotheses. This bacterium's soil habitat is spatially and temporally complex, which hinders the evaluation of the motility function of each flagellar system in nature.

Consistent with previous studies in soft agar, where motility within the agar mesh is dominated by swimming~\cite{Althabegoiti2011}, our observations indicate that the presence of the two flagellar systems increases the swimming speed of the WT strain as compared to the mutants lacking either the lateral or the subpolar flagella. This points to a cooperative function of both flagellar systems, contrary to what has been observed in other dual-flagellated bacterial species, such as the marine bacterium \textit{V. alginolyticus}, where cells expressing both a lateral and a subpolar flagellar system swim slower than those expressing the subpolar flagellum only~\cite{Grognot2023}.

A faster swimming speed can indeed represent an evolutionary advantage for \textit{B. diazoefficiens} in the soil, as it must compete with other rhizobia for colonization of soil habitats and nodulation of legume host roots. In effect, although final effectiveness in nodulation depends not only on swimming speed but also on the spatial distribution of cells in the soil and rhizosphere colonization~\cite{Lopez-Garcia2002, Balda2024}, increased motility has been identified as one favourable trait in nodulation competition experiments~\cite{Lopez-Garcia2009, Althabegoiti2008, Althabegoiti2011}. However, our results in increasingly confined microchannels suggest that this advantage is limited to water-saturated pores of sufficiently large size, where higher motility can aid \textit{B. diazoefficiens} to spread in the soil, in agreement with previous experiments on the competition for nodulation~\cite{Althabegoiti2011}. Interestingly, the swimming behaviour inside narrow pores sized similarly to their body sizes, channels of few micrometres in width, appears to be dominated by the subpolar flagellum, as evidenced by the striking resemblance of all the measured motility parameters of the WT and the $\Delta$\textit{lafA} strains in the most microconfined geometry.

Despite previous observations indicating that the subpolar flagellum of \textit{B. diazoefficiens} is able to reverse its spinning direction~\cite{Kanbe2007, Althabegoiti2011}, our close inspection of cell trajectories revealed only RR changes of direction occurring by turning the whole cell body at \SI{180}{\degree}. Although our observations do not rule out the possibility of RRs by changing the rotation sense of the subpolar flagellum, such mode appears to be much less common than RRs by a full rotation of the bacterial body. This clearly contrasts the known model of \textit{V. alginolyticus}~\cite{Xie2011, Son2013} and \textit{C. crescentus}~\cite{Morse2016}. This way of changing swimming direction by turning the whole body orientation might be an adaptation to the simultaneous presence of lateral flagella in the swimming cell, which might preclude the RR motion if it depended solely on the spinning reversal of the subpolar flagellum.

\begin{figure*}[htb]
\centering
\includegraphics[width=0.8\textwidth]{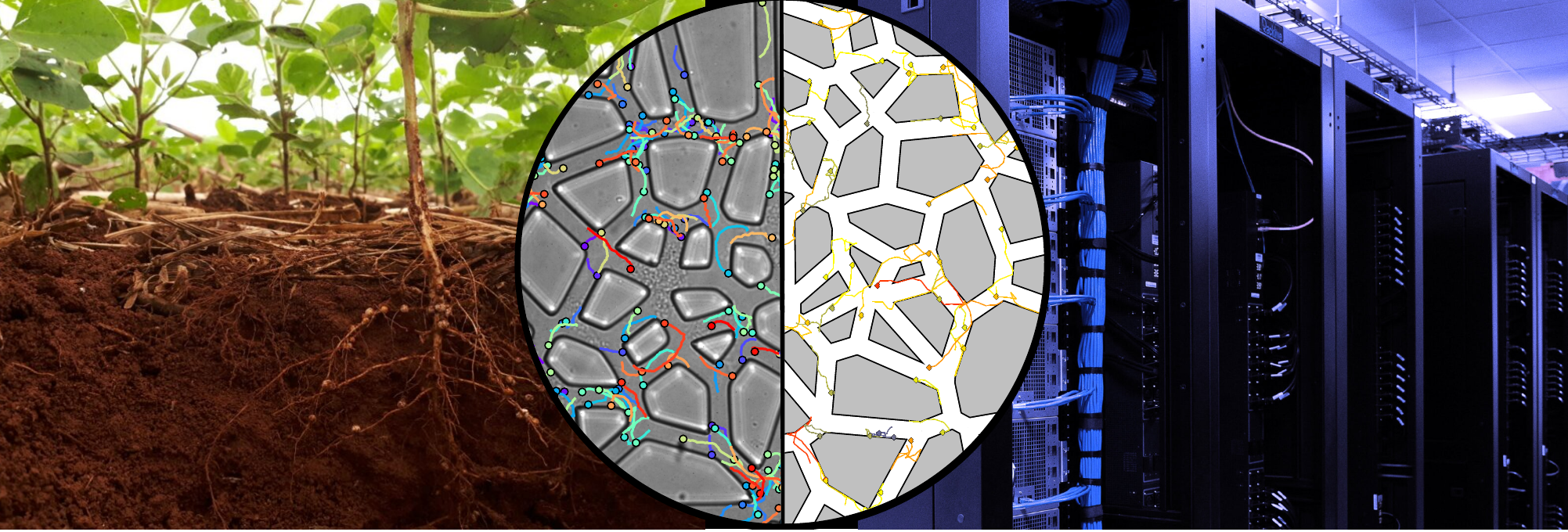}
\caption{\label{fig:Concluding} 
\textbf{Microfluidics devices for Studying \textit{B. diazoefficiens} Confined Swimming behaviour: Experimental vs. Theoretical approaches}. The central circle highlights the reached goals of this work: nice agreements between real-time microbial tracking in microfluidic devices and predictive theoretical models and calculations, utilizing high-performance computing.}
\end{figure*}

Swarming is another flagella-driven motility that should be considered to explain these results. Past swarming experiments performed with semisolid agar, however, have demonstrated that the mutant with lateral flagella only ($\Delta$\textit{fliC}) has superior performance than the one with the subpolar flagellum only ($\Delta$\textit{lafA}), which in turn is just marginally lower than the WT~\cite{Covelli2013}. However, experiments in soil pills indicated that the $\Delta$\textit{fliC} strain possesses lower motility than the $\Delta$\textit{lafA} or the WT, suggesting that swarming is not important in soil~\cite{Covelli2013}. Furthermore, swarming is a social motion that requires the association of various bacterial cells and the release of surfactants to lubricate the surface on which this motion is exerted~\cite{Verstraeten2008}. In our microfluidic devices, particularly in the narrowest channels microsized as their body sizes, association of various bacterial cells is not seen and almost not possible due to space constraints. Flagellar filaments may also act as adhesins to mediate cell adhesion to solid surfaces. Adhesion to glass--which is a similar surface to silicate soil particles--was stimulated by the subpolar flagellar adhesin FliC, but the WT has an adhesion similar to the $\Delta$\textit{fliC} or the non-flagellated mutants, leading us to hypothesize that LafA might preclude the adhesin activity of FliC avoiding premature adhesion, and thus favouring swimming near the solid boundaries of soil grains~\cite{Quelas2016}. This trend was confirmed in our experiments, where the motility was driven solely by swimming in the liquid medium, even along the solid boundaries, and where the $\Delta$\textit{lafA} and the WT had similar performances. Interestingly, the expression of lateral flagella, as well as the swimming speed of the $\Delta$\textit{fliC} mutant, were observed to increase when the viscosity of the medium in augmented through the addition of polyvinylpyrrolidone~\cite{Mengucci2020}. These observations support the idea that lateral flagella are relevant for motility of \textit{B. diazoefficiens} in complex and varying spatiotemporal conditions present in the soil, for example, in viscous layers, and highlight the need for observations of \textit{B. diazoefficiens} in non-Newtonian fluids, where lateral flagella could be advantageous~\cite{Grognot2023}.

Although several artificial models of soil have been proposed in the last decade~\cite{Downie2012, Stanley2016, Sharma2020}, our contribution includes reproducible geometrical details to mimic the real grain and pore sizes and distribution. However, although representing an advance thanks to the ability to directly observe the motility of soil bacteria in an environment that mimics their natural habitat, our microfluidic devices still represent a simplistic approach, lacking more realistic chemical, rheological and compositional complexity to simulate natural soil. In this respect, future endeavours should incorporate chemical signals and more complex porous media~\cite{deAnna2021, Scheidweiler2020, Bordoloi2022, Scheidweiler2024}, non-Newtonian fluids~\cite{Grognot2023}, organic matter~\cite{Yang2021a}, fluid flows~\cite{Scheidweiler2020} and different degrees of water deficit~\cite{Yang2021b}, in addition to geometrical constraints.

The experiments presented here reveal an unexpected decrease in swimming speed and an increasing similarity of the motility parameters of WT \textit{B. diazoefficiens} and its mutant strain lacking lateral flagella inside micro-confined water-saturated pores. This kind of experiment, however, cannot follow the long-term behaviour of bacterial populations and is spatially limited by the optical field of view of microscopy techniques. In this sense, simple, precise and efficient numerical simulations able to predict the behaviour of a large amount of soil bacteria with realistic experimentally measured parameters are valuable, to infer bacterial behaviour both at microscopic and larger scales. The proposed phenomenological model offers the opportunity to scale further accurate predictions of transport properties in much larger system sizes and time scales than those attainable with microfluidic devices, a key ingredient for applications at the macroscale.

\section*{CONCLUSIONS}

Our work combined multidisciplinary contributions, from microfabrication technologies, fluids and computational physics and genetic-agronomic background to understand better the behaviour of \textit{B. diazoefficiens} in the soil that might be of value for the development of a new generation of biofertilizers (summarized in Fig.~\ref{fig:Concluding}). So far, the motility performance of different strains could only be compared directly by microscopy in a simple homogeneous liquid medium or by measuring colony growth on agar plates, where the geometrical complexity and confinement naturally existing in real soils are absent. On the contrary, our biocompatible and transparent microfluidic devices allowed direct visualisation of bacteria behaviour in conditions mimicking soil granular and porous structure, providing valuable insights into the behaviour of this beneficial bacterium in controlled and reproducible experiments and numerical calculations.

\section*{METHODS}

\subsection*{Microfluidic devices}

Microfluidic devices were fabricated using standard lithography and soft lithography methods. Templates were fabricated using optical lithography on 2'' silicon wafers using SU-8 (Gersteltec Sarl) with a maskless laser writer (MLA100, Heidelberg Instruments). Polydimethylsiloxane (PDMS, Sylgard 184, Dow Corning) replicas were obtained from the moulds. Once cured, the PDMS device was cut with a scalpel and carefully detached from the moulds. Inlet and outlet holes were pierced in the PDMS blocks with a \SI{1.5}{\milli\meter}-diameter biopsy punch, after which both the PDMS block and a microscope glass slide were exposed to an air plasma for one minute and sealed together. The plasma bonding produced an irreversible covalent bonding between the PDMS grains and the glass~\cite{Xiong2014}, making the channels the only accessible region for the bacteria.

The devices comprised an inoculation chamber, a narrow, randomly oriented channel network region, and an exit chamber. The network was designed in Matlab as a Voronoi tessellation, obtained from $N = 3000$ seed points within a region of $A = 2 \times 1$ (arbitrary units)~\cite{Wu2012, Ponce2020}. To generate a disordered network, these points were required to be at least $0.3 d_0$ units apart, where $d_0 = \sqrt{2 A/(N\sqrt{3})}$ would be the distance between the seed points in a completely regular network (a honeycomb network), but other than that they were randomly generated. The edges of the Voronoi polygons were then given a fixed width, and the network was re-scaled to real dimensions. The same design was used for the three networks, which differed in their scaling factor and final height (defined in the lithographic step). The CAD file used for the maskless lithography is available as Supplementary Data. The final dimensions (height $h_i$, channel width $w_i$ and overall area of the network $L_i \times W_i$, $i=1,2,3$) of the devices were (i) $h_1=\SI{25}{\micro\meter}, w_1=\SI{20}{\micro\meter}$ and area $4\times 2$~\si{\milli\meter^2}, (ii) $h_2=\SI{25}{\micro\meter}, w_2=\SI{10}{\micro\meter}$ and area $2\times 1$~\si{\milli\meter^2}, and (iii) $h_3=\SI{15}{\micro\meter}, w_3=\SI{5}{\micro\meter}$ and area $1\times 0.5$~\si{\milli\meter^2}.

\subsection*{Bacterial Strains and Culture}

\textit{B. diazoefficiens} USDA 110 (obtained from the United States Department of Agriculture, Beltsville) is the type strain of the species. The derived mutants $\Delta$\textit{fliC} and $\Delta$\textit{lafA}, which are devoid of subpolar or lateral flagella respectively, were described elsewhere~\cite{Althabegoiti2011}.

Bacteria were grown in HM salts medium~\cite{Cole1973}, supplemented with \SI{1}{\gram/\liter} yeast extract and \SI{5}{\gram/\liter} L-arabinose (HMY-Ara), as described~\cite{Quelas2016}. For routine use, bacteria stocks were kept at \SI{4}{\celsius} in yeast extract mannitol-agar (\SI{1.5}{\percent})~\cite{Vincent1970}, supplemented with chloramphenicol \SI{20}{\milli\gram/\liter} plus kanamycin \SI{200}{\milli\gram/\liter} ($\Delta$\textit{fliC}) or streptomycin \SI{400}{\milli\gram/\liter} ($\Delta$\textit{lafA}). Cultures in agar plates were renewed every 3 months from frozen stocks kept at \SI{-80}{\celsius} in culture medium HMY-Ara supplemented with 20\% v/v glycerol. Liquid cultures were initiated from a single colony in HMY-Ara and grown for seven days at \SI{28}{\celsius} and rotary shaking at \SI{180}{\rpm} until the optical density at \SI{600}{\nano\meter} (OD$_{600}$) reached OD$_{600} = 3.0 \pm 0.1$. A 1:100 dilution into HMY-Ara was then grown for two days at the same temperature and agitation conditions until OD$_{600} = 1 \pm 0.1$. A 1:100 dilution into HM supplemented with L-arabinose \SI{5}{\gram/\liter} (HM-Ara) was then left at \SI{28}{\degree} without agitation for a minimum of \SI{6}{\hour} before chamber inoculation.

At least three different cultures for each strain, starting from different colonies on different days (biological replicas), were used to ensure the reproducibility of the experiments.

\subsection*{Inoculation and data acquisition}

Microfluidic devices were filled with HM-Ara medium before use to eliminate air bubbles. Bacteria were inoculated into the devices at constant flow rates, between \SI{1.6}{\nano\liter/\second} and \SI{30}{\nano\liter/\second}, using \SI{1}{\milli\liter} glass syringes (Hamilton Co.), a syringe pump (neMESYS Base 120 and low-pressure module V2, Cetoni) and Tygon tubes (0.02'' ID x 0.06'' OD Microbore Transfer Tubing, Masterflex) until bacteria had completely entered the device. Tubing was then cut near the syringe needle, and the tubing ends were introduced into a reservoir with HM-Ara. After an equilibration time required to relax any remaining flow, bacteria were observed in a bright field using an inverted microscope (TS100 LED, Nikon) equipped with a 40X microscope objective and a CMOS camera (Zyla 4.2, Andor or Mini AX100, Photron). The spatial resolution of the optical system was \SI{6.24}{\pixel/\micro\meter} with the Andor camera and \SI{3.06}{\pixel/\micro\meter} with the Photron camera. The effective depth of field was determined to be \SI{4}{\micro\meter} by fixing the focus on bacteria attached to the glass surface and slowly moving the objective until the attached bacteria were no longer detectable. It is important to remark that the focus depth is merely four times the bacteria dimensions, then they can cross it just in \SI{0.8}{\second} if they swim at \SI{22}{\micro\meter/\second}. Several \SI{60}{\second} image sequences were recorded at 50 fps for later analysis, focusing on different sites of the network but also at the inoculation and exit chambers.

\subsection*{Bacteria tracking}

The bacteria trajectories over time were obtained through image analysis of the experimental videos. The positions were extracted using the open software {\it Biotracker}, developed within our group of collaborators in FaMAF-National University of Córdoba~\cite{Sanchez2016, Reyes2017, biotracker}. BioTracker is an open-source computer vision framework designed for visual cell tracking. The software is suitable for a variety of image recording conditions and offers a number of different tracking algorithms. Several well-known particle tracking software available are built generically~\cite{Chenouard2014} and may not be optimized for fast microswimmers or small cells that are hard to detect. To address this issue, we developed BioTracker to follow fast and sub-micron-sized microswimmers, with specific motility features such as direction changes along their paths or complex 3D oscillations of sperm cells~\cite{Bettera2020}. When the software started to be developed~\cite{Sanchez2016}, tracking fast-moving bacteria with high precision without confusing the tracks' projections and crossings was a significant challenge. Despite this, the software was able to deliver better results than the 14 methods compared in the article by Chenouard \textit{et. al}, published in Nature Methods 2014~\cite{Chenouard2014}. The software utilizes Kalman filter estimators in its linker module, which contributed to its quality in tracking fast-moving bacteria.

The BioTracker software offers several advantages in analysing videos of long duration with precision. It demonstrates stability in performance over various metrics and scenarios, low computational cost, and robustness. Specifically built to track microswimmers with rapid and unpredictable movements, including fast and abrupt runs and tumbles, BioTracker automatically counts and characterizes such events and their angles in a dedicated module. Its quality is attributed to its detector module and the qualified post-acquisition analysis of subtle details of the tracks, such as small changes of directions or run and reverse events, which are extensively used in the research presented here. The software provides a user-friendly interface and supports the implementation of new problem-specific tracking algorithms, allowing the scientific community to accelerate their research and focus on developing new vision algorithms.

A change of direction is detected in a given position if there is a change in the swimming orientation $n_b$ body lengths behind and ahead of the current position, where the user chooses the number $n_b$. For a correct implementation, it is fundamental to characterise the cell body sizes well. Moreover, the population can be heterogeneous, and BioTracker takes that into account. A lower threshold for the CHD angle was defined at \SI{15}{\degree} to distinguish it from the usual Brownian motion.

In this work, we were able to detect and track thousands of soil minute bacteria simultaneously and precisely extract all their motility parameters, not just their main speed, using BioTracker. By fine-tuning the software options, we were able to achieve precise confined motility measurements of {\it B. diazoefficiens}, a bacterium that is extremely small and hard to detect and even harder to follow in-plane and to detect their run-and-reverses. These events occur very fast in distances of the order of their body sizes. 

\subsection*{Statistics and Reproducibility}

In principle, three aliquots from a given culture were used to inoculate the three networks on the same day. However, when statistics were not deemed enough for a given network and/or strain, additional measurements were obtained with a different culture in a specific network. Hundreds to thousands of tracks were obtained for each replicate, and all relevant distributions were obtained from them. The similarity of the obtained distributions confirmed the reproducibility of the experimental conditions.

\subsection*{Model and Simulations}

We simulated the behaviour of soil bacteria diluted in an aqueous medium and confined in mimicking soil microdevices at low Reynolds numbers~\cite{Purcell77}. This system is characterized predominantly by viscous effects acting on cells. This means that forces act instantaneously, and the speed of motion of the microswimmers is directly proportional to their motor force of self-propulsion. Consequently, the important velocity parameter captures in a very simple and amazing manner both the biological complexity of the bacteria motors and all physical forces involved in their motions. We model and simulate $N_b$  self-propelled bacteria, reproducing exactly the concentration of cells inoculated in each experiment, using their corresponding measured speeds $v$ and directions of motion associated with its flagellated motor forces (the PDFs reported in previous figures for each geometry were used). Each strain has its unique and specific swimming strategy, which is incorporated into the model thanks to the use of the velocity statistic measured for both strains. Our aim was to develop a simple yet extremely accurate model that can efficiently simulate large populations of $N_b$ microswimmers (thousands of cells) interacting among them and very often interacting with the walls of the many grains that mimic soil geometries. As bacteria swim inside the microfluidic devices, surfaces dominate, and interactions with walls significantly impact the simulation's CPU time. To achieve realism, we utilized measured data for the motility parameters of each strain (Figs.~\ref{fig:MeanSpeed} and \ref{fig:AngleDistributions}) without approximating biological parameters in the calculations. 

The phenomenological model we proposed for the confined bacteria behaviour shares similarities with previous useful proposals for other bacteria swimming in microdevices or porous media~\cite{Berdakin2013a, Berdakin2013b, Montagna2018}. The dominant forces we considered are: 1. Motor force, $\textbf{F}^{m}$. 2. Bacteria interactions with walls, $\textbf{F}^{bw}$. 3. Fluid damping force proportional to speed and $-\gamma$, the damping constant in an aqueous medium. Although the system is diluted, we also accounted for soft interactions among bacteria, $\textbf{F}^{bb}$, for completeness.

Here, we adapted and optimized previous models to reduce computing time, accommodating the geometry of microfluidic devices and the speed and specific change of direction distributions of \textit{B. diazoefficiens}, as run-and-reverse movements, which are crucial. The initial swimmer's positions were randomly distributed across the microfluidic device (see Fig.~\ref{fig:Simulations}). The initial speeds followed the same measured speed distributions (Fig.~\ref{fig:MeanSpeed}), and it is imposed to be constant along the simulated time. From this starting point, we have a predictive set of  2$N_b$ equations of motion. The overdamped motion equation (viscous motion) for each soil bacterium, positioned at $\textbf{r}$ and resulting from the cancellation of the sum of forces acting on its body, is as follows: $\gamma d\textbf{r}/dt = \textbf{F}^m+\textbf{F}^{bw} +\textbf{F}^{bb}$. In addition, a coupled time-evolution equation for the velocity direction was added, where the measured change of directions was imposed~\cite{Berdakin2013a, Berdakin2013b, Montagna2018} with its proper statistics and probability.

The bacteria swimming in very flat microdevices were modelled in 2D as disks with a radius of $R_b=$ \SI{0.5}{\micro\meter} because the average measured diameter in our experiments is \SI{1}{\micro\meter}. We used real measured parameters to propose the equations of motion for positions and velocity direction, as described in Refs.~\cite{Berdakin2013a, Berdakin2013b, Montagna2018}. The equations of motion are solved using Langevin dynamics methods. We introduced the exact change of direction distribution measured, as shown in Fig.~\ref{fig:AngleDistributions}, using the rejection method of Von Neumann to reproduce the distributions exactly, rather than relying on approximations commonly used in the literature and in our previous contributions. We also optimized the codes to speed up calculations of bacteria-grain interactions, resulting in a \SI{50}{\percent} improvement in performance compared to previous codes (optimization modules are available; contact the corresponding authors). As a result, the confined trajectories at all times were obtained in our variety of microfluidic devices with different geometries and channel widths (Fig.~\ref{fig:Simulations}) in a good experimental-numerical agreement.

\section*{ACKNOWLEDGEMENTS}

This work was partially funded by ANID -- Millennium Science Initiative Program-NCN19\_170, ANID Fondecyt grant 1210634 (Chile), Secyt-UNC for travel Chile-Argentina and Project Consolidar 33620180101258CB,  CONICET with PIP2023-11220220100509CO and FONCyT with PICT-2020-SERIEA-0293 (Argentina). Fabrication of microfluidic devices was possible thanks to ANID Fondequip grants EQM140055 and EQM180009 (Chile). Simulations were done in the computational facilities of Universidad Nacional de Córdoba, Argentina. MPM acknowledges funding from ANID Fondecyt postdoctoral grant No. 3190637 (Chile). JPCM acknowledges funding from ANID Beca de Magíster Nacional No. 22221639 (Chile). Experimental assistance from Francisca Paredes Oltra and a careful reading of the manuscript from Prof. Pedro Pury and Prof. Adolfo J. Banchio from UNC-Argentina, as well as useful discussions, are greatly appreciated.\\

\section*{AUTHOR CONTRIBUTIONS STATEMENT}

VIM, MLC, and AL conceived the work. MPM and JPCM conducted the experiments. NG, MPM, JPCM, and SM analyzed experimental data. VIM developed the model and offered Biotracker. NG, SM, and VIM performed simulations. MLC, VIM and AL wrote the manuscript. All authors reviewed the manuscript.

\section*{COMPETING INTERESTS}

The authors declare no competing interests.

\section*{DATA AVAILABILITY}

All main data generated or analyzed during this study were deposited in the Data Repository at Universidad de Chile, \url{https://doi.org/10.34691/UCHILE/V9SGXW}.

\section*{CODE AVAILABILITY}

All details on codes developed are available from the corresponding author upon reasonable request. Biotracker is available on GitHub \cite{biotracker}.

\section*{ADDITIONAL INFORMATION}

 The online version contains supplementary material available at...
 
Correspondence and requests for materials should be addressed to María Luisa Cordero or V. I. Marconi.


\bibliography{main}

\end{document}